\documentclass[twocolumn,showpacs,preprintnumbers,prl]{revtex4}
\usepackage{graphicx,epsfig}

\def\simgt{\rlap{\lower 3.5 pt \hbox{$\mathchar \sim$}} \raise 1pt \hbox {$>$}}
\def\simlt{\rlap{\lower 3.5 pt \hbox{$\mathchar \sim$}} \raise 1pt \hbox {$<$}}

\begin{document}


\title{
\boldmath Recent progress in $B$ physics \unboldmath}
\author{Dongsheng Du%
        \thanks{Invited talk presented at the $10^{\rm th}$ Asia
         Pacific Physics Conference (APPC10), August 21-24, 2007, Pohang, Korea.}}

\affiliation{ Institute of High Energy Physics,
 Beijing 100049, China}

\date{September 6, 2007}

\begin{abstract}
\noindent We firstly address the recent efforts on calculations of
the next-to-leading order corrections to the color-suppressed tree
amplitude in QCD factorization method which may be essential to
solve the puzzles in $B\to \pi\pi$ and $\pi K$ decays. Then we
discuss the polarization puzzles in $B\to \phi K^*$ and $\rho
K^*$. The impacts of the newly measured $B_s-\bar B_s$ mixing and
$B^+\to \tau^+\nu$ on the CKM unitarity triangle global fitting
are mentioned. We also briefly review the recent measurements of
the new resonances at BaBar and Belle. Finally, some new results
from hadron colliders, especially the $b$-flavored hadron spectra,
are discussed.

\end{abstract}

\maketitle

\section{Introduction}

\noindent The study of $B$ meson decays plays an important role in
determining the CP-violating parameters in the Standard Model (SM)
and discovering new physics in the flavor-changing processes. In
particular, $B$ non-leptonic two-body decays provide an abundant
sources of information about the CKM matrix. For example, the most
promising measurement of $\sin 2\beta$ ($\alpha$,$\beta$ and
$\gamma$ are three angles in the unitarity tri-angle) is from
measurement of the time-dependent CP-asymmetry in $B^0\to J/\Psi
K_S$. $B\to \pi\pi$, $\rho\pi$ and $\rho\rho$ are also very
important for determining $\sin 2\alpha$. However, the theoretical
calculation of these hadronic decays suffers from the complicated
strong interactions which compromise the precision of the
determination of the CKM matrix elements from the experimental data.
Thus the higher order calculation of such decays are essential for a
better understanding of the CP violations.

Besides the non-leptonic decays, the new measurements on $B_s-\bar
B_s$ mixing and $B^+\to \tau^+\nu$ have a great impact on
constraining the unitarity tri-angle. Both of two decay modes are
also sensitive to new physics.

 The $B$ decays also offer us a good place to study the
strong interaction dynamics in heavy flavor systems. In the abundant
decay products of $B$ mesons, experimentalists observed a lot of new
hadron resonances at BaBar and Belle. They are mainly excited (or
exotic) charmed mesons and charmonium states.

 The excited $B$-mesons and $b$-baryons can be only studied at the hadron
 colliders. We will briefly review the spectrum of the excited $B$ mesons
 and $b$-flavored baryons from the measurements at TEVATRON.

\section{$B\to \pi\pi$ and $\pi K$ puzzles}

$B\to \pi \pi$ and $\pi K$ are the most popularly studied two-body
decay modes in $B$ physics in addition to $B\to J/\Psi K_S$. The big
experimental achievement is that the direct CP asymmetries in these
decays have been observed. The latest world averages are\cite{HFAG}
\begin{eqnarray}
A_{CP}(\pi^+\pi^-)|_{\rm exp.}&=&0.38\pm
0.07,\nonumber\\
A_{CP}(\pi^+K^-)|_{\rm exp.}&=&-0.095\pm 0.013,
\end{eqnarray}
both of which are $5\sigma$ away from zero.

 $B\to \pi\pi$ are dominated by tree amplitudes.
Due to the isospin symmetry, the decay amplitudes of tree $B\to
\pi\pi$ modes can be parameterized graphically as the following
\begin{eqnarray}
{\cal A}(\pi^+\pi^-)&=& T e^{-i\gamma}+P,\nonumber \\
\sqrt{2}{\cal A}(\pi^0\pi^-)&=& (T +C)e^{-i\gamma},\\
{\cal A}(\pi^0\pi^0)&=& -C e^{-i\gamma}+P,\nonumber
\end{eqnarray}
where $T$, $C$ and $P$ stand for the color-allowed, color-suppressed
tree amplitude and penguin amplitude respectively. According to the
naive factorization, the ratios $|C/T|$ and $|P/T|$ are expected to
be small. This leads to that the
 ${\rm Br}(\pi^0\pi^-)$ is almost half of  ${\rm
 Br}(\pi^+\pi^-)$, and  ${\rm Br}(\pi^0\pi^0)$ is very small.
 However, this expectation is strongly against the experimental
 data\cite{HFAG}
 \begin{eqnarray}
10^6\,{\rm Br}(\pi^+\pi^-)_{\rm exp.}&=& 5.16\pm 0.22,\nonumber \\
10^6\,{\rm Br}(\pi^0\pi^-)_{\rm exp.}&=& 5.7\pm 0.4, \\
10^6\,{\rm Br}(\pi^0\pi^0)_{\rm exp.}&=& 1.31\pm 0.21,\nonumber
 \end{eqnarray}
which requires $|C/T|\simeq 0.7$. It means that the
color-suppression is not valid any more.

For $B\to \pi K$ decays, the similar graphical parameterization can
be written as
\begin{eqnarray}
\label{bpikamp}
A(\pi^-\bar K^{0})&=&P^\prime,\nonumber \\
\sqrt{2}
A(\pi^0K^{-})&=&[P^\prime+P^{EW}]+e^{-i\gamma}[T^\prime+C^\prime],\nonumber
\\
A(\pi^+K^{-})&=& P^\prime+e^{-i \gamma}T^\prime,
\\
-\sqrt{2}A(\pi^0\bar
K^{0})&=&[P^\prime-P^{EW}]-e^{-i\gamma}C^\prime,\nonumber
\end{eqnarray}
in which penguin amplitude $P^\prime$ dominates comparing with the
color-allowed tree amplitude $T^\prime$, color-suppressed tree
amplitude $C^\prime$ and the electro-weak penguin $P^{EW}$.
$A_{CP}(\pi^+K^-)\simeq A_{CP}(\pi^0 K^-)$ is expected if
$|C^\prime/T^\prime|$ and $|P^{EW}/P^\prime|$ are small. However,
the recent experimental data shows $A_{CP}(\pi^0
K^-)=0.047\pm0.026$. It requires either the enhancement in
$P^{EW}$ or $C^\prime$. The large $P^{EW}$ scenario
\cite{Yoshikawa:2003hb} seems to be going away with the new
experimental measurement of ${\rm BR}(\pi K)$.
$|C^\prime/T^\prime|\simeq 1.1$ is needed to meet the data. So
similar to the situation in $B\to \pi\pi$, the color-suppression
for $C^{(\prime)}$ is not valid any more.

In QCD factorization, these color-suppressed amplitudes are
related to the QCD coefficient $\alpha_2(M_1 M_2)$. For
illustration\cite{Beneke:2003zv},
\begin{eqnarray}
\alpha_2(\pi\pi) &=& 0.17_{ rm LO} - [0.17+0.08i\,]_{V_2}
\\&+& \left\{\begin{array}{lc} [0.18]_{\rm LOHSI} & \qquad
\mbox{(default)}
\\
{[0.46]_{\rm LOHSI}} & \qquad \mbox{(S4)}
\end{array}
\right. \nonumber
\end{eqnarray}
The accidental cancellation between the leading order (LO) and
next-to-leading order (NLO) vertex corrections ($V_2$) makes the
hard spectator interaction (HSI) very important. The NLO corrections
to the HSI are recently studied by Beneke, Jager and
Yang\cite{Beneke:2005vv,Beneke:2006mk,Beneke:2005gs}. In their
papers, the corrections from the two scale regions are encoded into
the jet function and hard coefficient respectively, both of which
enhance the color-suppressed amplitude effectively.

 In Table \ref{table1}, the predictions in QCDF with
NLO HSI in a certain parameter setting ($G_4$) is shown. The
agreement between the prediction and experimental data is very
good except for the direct CP asymmetries $A_{CP}$. It means that
the strong phases still need further study. Recently, the efforts
towards the NLO corrections to the imaginary part of the amplitude
has started. In \cite{Bell:2007tv}, the next-next-to-leading order
vertex corrections has been considered.

\begin{table}
\begin{tabular}{c|c|c|c|c|c}\hline\hline
${\rm Br}\times 10^6$ & $G_4$ & Exp.
& $A_{CP}$ & $G_4$ & Exp.\\
\hline $\pi^0\pi^-$ & $5.6$ & $5.7\pm 0.4$ & $\pi^0\pi^-$ & $0.00$
& $0.04\pm 0.05$
\\ \hline
$\pi^+\pi^-$ & $5.7$ & $5.16\pm 0.22$ & $\pi^+\pi^-$ & $0.04$ &
$0.38\pm 0.07$
\\ \hline
$\pi^0\pi^0$ & $0.81$ & $1.31\pm 0.21$ &$\pi^0\pi^0$ & $-0.38$ &
$0.36\pm 0.33$
\\ \hline
$\pi^-\bar{K}^0$ & $22.6$ & $23.1\pm 1.0$ & $\pi^-\bar{K}^0$ &
$0.00$ & $0.008\pm 0.0025$
\\ \hline
$\pi^0 K^-$ & $12.9$ & $12.8\pm 0.6$ & $\pi^0 K^-$ & $-0.05$ &
$0.047\pm 0.026$
\\ \hline
$\pi^+ K^-$ & $20.6$ & $19.4\pm 0.6$ & $\pi^+ K^-$ & $-0.02$ &
$-0.095\pm 0.013$
\\ \hline
$\pi^0\bar{K}^0$ & $9.1$ & $10.0\pm 0.6$ & $\pi^0\bar{K}^0$ &
$0.04$ & $-0.12\pm 0.11$
\\ \hline
\hline
\end{tabular}
\caption{Predictions in QCDF with NLO HSI vs. experimental
data.}\label{table1}
\end{table}

\section{$B\to VV$ and polarization puzzles}
$B$ decays to two light vector mesons offers more abundant
observables than $B\to PP$ and $PV$ since the final vectors can be
polarized both longitudinally and transversely. The fact of the
$V-A$ dominance in the Standard Model (SM) shows the hierarchy of
the helicity amplitudes for $B\to VV$
\begin{equation}
A_0:A_-:A_+ = 1 : \frac{\Lambda}{m_b} :
\left(\frac{\Lambda}{m_b}\right)^{\!2} \label{hierarchy}
\end{equation}
with simple estimation by the naive factorization. This argument
leads to the expectation that $B\to VV$ is dominated by
longitudinal polarization. However, experimentally, such hierarchy
is obeyed in the tree-dominated decays ($B\to \rho\rho$ and
$\omega \rho$), but violated in penguin dominated $B\to \phi K^*$
system\cite{Aubert:2003mm}. This attracts a lot attentions from
the theorists. Many solutions are offered on the table, such as
new physics (scalar or tensor coupling), final state interaction,
charming penguin, the form-factor tuning, large penguin
annihilation etc\cite{PhiKstar}.

The more puzzling case happens in $B\to \rho K^*$ system which are
penguin-dominated, in which both transverse polarization is enhanced
in $B^-\to \rho^- \bar K^{*0}$ but suppressed in $\bar B^-\to \rho^0
K^{*-}$\cite{RhoKstar}.

Analogue to $B\to PP$ and $PV$, QCD factorization formula for $B\to
VV$ reads\cite{Beneke:2006hg}
\begin{eqnarray}
 &&\langle V_{1,h} V_{2,h}|Q_i|\bar B\rangle
   =  F^{B\to V_1,h}\,T_{i}^{I,h} * f_{V_2}^{h}\Phi_{V_2}^{h} \nonumber\\
   &
    +& T_i^{II,h} * f_B\Phi_B * f_{V_1}\Phi_{V_1} *
    f_{V_2}^h\Phi_{V_2}^h+{\cal O}(1/m_b)\,.
\end{eqnarray}
Also similar to $B\to PP$ and $PV$, the annihilation contribution
brings the large uncertainties to the QCD factorization prediction
in $B\to VV$ as well due to the severe endpoint singularity.
However, the penguin weak annihilation in longitudinal helicity
amplitude is predicted to be small and with small uncertainty,
perhaps due to an accidental cancellation; in negative-helicity
amplitude, the penguin weak annihilation could be very large. This
leads to two consequences:

1) For the tree-dominated decays which are dominated by the
longitudinal polarization, the penguin amplitude gives small
contribution to the branching ratios. As a result, we could have a
better determination of $\sin 2\alpha$ or $\alpha$ from $B\to
\rho\rho$ than $B\to \pi\pi$ and $\pi \rho$.

2) For the penguin dominated decays, the negative-helicity
amplitude could be (need not to be) large. This can be an solution
for the $B\to \phi K^*$ polarization puzzle. However, in this
case, QCD factorization loses its predictive power.

\begin{figure}[t]
\epsfig{file=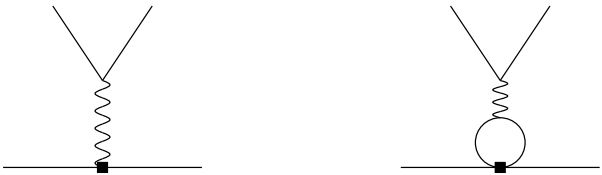,width=8cm,angle=0}
 \caption{\label{fig1} Leading contributions to $\Delta
\alpha_{3,\rm EW}^{p\mp}(V_1 V_2)$ defined in the text.}
\end{figure}

The decay amplitudes for $B\to \rho K^*$ system could be
parameterized graphically as Eq.(\ref{bpikamp}) except for each
amplitude replaced by the one noted by helicity. To explain the
polarization fraction in $B\to \rho K^*$ system, we need large
electroweak penguin for negative-helicity amplitude. Such
enhancement can be obtained if consider the electromagnetic dipole
operator in the effective weak Hamiltonian
\begin{eqnarray}
{\cal H}_{\rm eff} &=& \frac{G_F}{\sqrt{2}} \sum_{p=u,c}
\lambda_p^{(D)} \sum_{a=-,+} C_{7\gamma}^a Q_{7\gamma}^a
+\ldots,\\
Q_{7\gamma}^\mp &=& -\frac{e\bar{m}_b}{8 \pi^2} \,\bar
D\sigma_{\mu\nu}(1\pm \gamma_5)F^{\mu\nu} b, \label{o7}
\end{eqnarray}
where $\lambda_p^{(D)}=V_{pb}V_{pD}^\ast$. This effect is
neglected in any calculation of $B\to VV$ before
\cite{Beneke:2005we}. In Figure \ref{fig1}, the small virtuality
of the photon can not be cancelled when the final vector $V_2$ is
transversely polarized. This results in the enhancement of the
electorweak penguin in negative-helicity amplitude
\begin{equation}
\Delta P^{\rm EW}(V_1 V_2) \propto  - \frac{2\alpha_{\rm
em}}{3\pi}\,C_{7\gamma,\rm eff}^- \frac{m_B\bar{m}_b}{m_{V_2}^2}
.\label{dal3ew}
\end{equation}
In Table \ref{table2}, we can see this effect obviously where the
penguin amplitudes are extracted from the experimental data.

In Table \ref{table3}, the predictions for $B\to \phi K^*$ and $\rho
K^*$ by QCD factorization are listed. ("$\hat\alpha^{c-}_4$ from
data" means the penguin of negative-helicity amplitude is extracted
from the experiments.) One can see that the predictions agree with
the experimental data very well. It reproduces the polarization
pattern of the penguin-dominated $B\to VV$ decays as well.
\begin{table}
 \begin{tabular}{lccc}
   \hline\hline
       $B^-\to K^{*-}\rho^0$ &
         with $\Delta P^{EW}$ & without $\Delta P^{EW}$ & experiment
        \\\hline
    ${\rm BrAv}/10^{-6}$
       & $4.5$
       & $5.4$
       & $<6.1$
        \\
    $f_L$ / \%
       & $84$
       & $70$
       & $96^{+6}_{-16}$
        \\\hline\hline
  \end{tabular}
\caption{The enhanced electroweak penguin in $B\to
VV$.}\label{table2}
\end{table}

\begin{table}[b]
  \begin{tabular}{lllll}\hline\hline

    \multicolumn{2}{l}{Observable} &
    \multicolumn{2}{l}{Theory} &
    Experiment
    \\\hline

    \multicolumn{2}{l}{}&
    default &
    $\hat{\alpha}_4^{c-}$ from data &
    \\\hline
    ${\rm BrAv} / 10^{-6}$
      & $\phi K^{*-}$
        & $10.1^{+0.5}_{-0.5}{}^{+12.2}_{-7.1}$
        & $10.4^{+0.5}_{-0.5}{}^{+5.2}_{-3.9}$
        & $9.7\pm 1.5$
        \\\hline
      & $\phi\bar{K}^{*0}$
        & $\phantom{1}9.3^{+0.5}_{-0.5}{}^{+11.4}_{-6.5}$
        & $\phantom{1}9.6^{+0.5}_{-0.5}{}^{+4.7}_{-3.6}$
        & $9.5\pm 0.8$
        \\\hline

     &  $\bar{K}^{*0} \rho^-$
        & $5.9^{+0.3}_{-0.3}{}^{+6.9}_{-3.7}$
        & $5.8^{+0.3}_{-0.3}{}^{+3.1}_{-1.9}$
        & $9.2\pm 1.5$
        \\\hline

    &   $K^{*-} \rho^0$
        & $4.5^{+1.5}_{-1.3}{}^{+3.0}_{-1.4}$
        & $4.5^{+1.5}_{-1.3}{}^{+1.8}_{-1.0}$
        & $< 6.1$
        \\\hline

    $f_L / \%$
      & $\phi K^{*-}$
        & $45^{+0}_{-0}{}^{+58}_{-36}$
        & $44^{+0}_{-0}{}^{+23}_{-23}$
        & $50\pm 7$
        \\\hline
      & $\phi\bar{K}^{*0}$
        & $44^{+0}_{-0}{}^{+59}_{-36}$
        & $43^{+0}_{-0}{}^{+23}_{-23}$
        & $49\pm 3$
        \\\hline

     &  $\bar{K}^{*0} \rho^-$
        & $56^{+0}_{-0}{}^{+48}_{-30}$
        & $57^{+0}_{-0}{}^{+21}_{-18}$
        & $48.0\pm 8.0$
        \\\hline
    &   $K^{*-} \rho^0$
        & $84^{+2}_{-3}{}^{+16}_{-25}$
        & $85^{+2}_{-3}{}^{+9}_{-11}$
        & $96^{+6}_{-16}$
        \\\hline\hline
  \end{tabular}
  \caption{QCD factorization predictions for
  $B\to \phi K^*$ and $\rho K^*$ vs. experimental data.}
  \label{table3}
  \end{table}

\section{$B_s-\bar B_s$ mixing and $B^+\to\tau^+\nu$}
CDF and D0 have measured the $B_s-\bar B_s$ oscillation frequency
\cite{Abulencia:2006mq}. We get the average of
\begin{eqnarray}
\Delta m_s\,=\,(17.77\pm 0.10\pm 0.09){\rm ps}^{-1}\nonumber
\end{eqnarray}
Comparing with $B_d-\bar{B}_d$ mixing\cite{HFAG}
\begin{eqnarray}
\Delta m_d&=&(0.507\pm 0.004){\rm ps}^{-1},\nonumber
 \end{eqnarray}
we can set the bound for
 \begin{eqnarray}
|V_{td}/V_{ts}|\,=\,0.206\pm 0.008.\nonumber
\end{eqnarray}
Consequently,
\begin{eqnarray}
\gamma\,=\,(66\pm 6)^\circ.\nonumber
\end{eqnarray}

Belle has declared the evidence for purely leptonic decays $B^-\to
\tau^- \bar{\nu}$\cite{Ikado:2006un}. Combined with BaBar's
bound\cite{:2007xj}, the HFAG average is given\cite{HFAG}.
\begin{eqnarray}
{\rm Br}(\tau^-\bar{\nu})&=&\left\{
\begin{array}{ll}
(88\pm 68\pm 11)\times 10^{-6} & {\rm~~~BaBar}\\
(176^{+56+46}_{-49-51})\times 10^{-6} &{\rm~~~Belle}\\
(132\pm 49)\times 10^{-6} &{\rm~~~HFAG~ Average}
\end{array}
 \right.\nonumber
\end{eqnarray}

\begin{figure}[t]
\epsfig{file=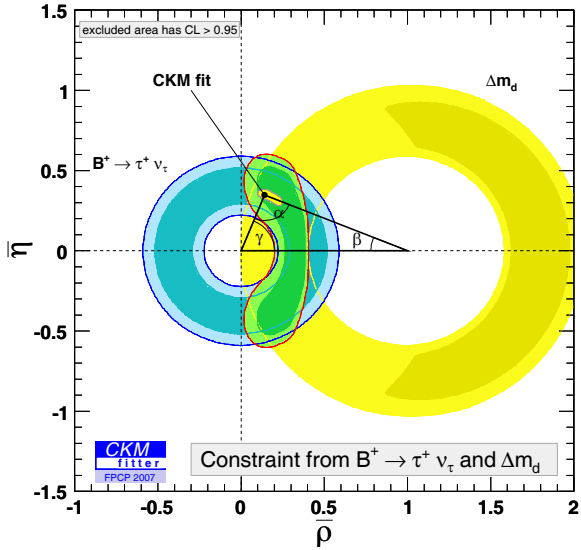,width=8cm,angle=0} \caption{Constraints
for $\bar \rho-\bar \eta$ plane from $B^+\to
\tau^+\nu$.}\label{fig2}
\end{figure}

\begin{figure}
\epsfig{file=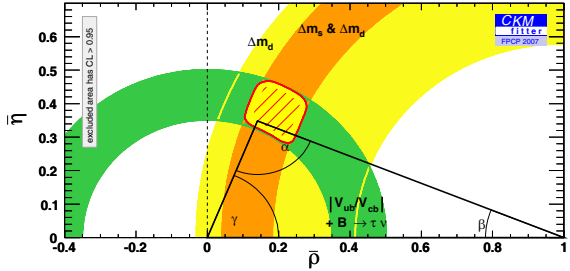,width=8cm,angle=0}
\caption{Combined constraints for $\bar \rho-\bar \eta$ plane from
$B^+ \to \tau^+\nu$ and $B_s-\bar B_s$ mixing.}\label{fig3}
\end{figure}

In the SM, the branching ratio of $B^-\to \tau^-\bar \nu$ can be
written as
\begin{eqnarray}
{\rm BR}(B^-\to \tau^-\bar \nu)_{\rm SM}&=&\frac{G_F^2 m_B
m_\tau^2}{8\pi}\left (1-\frac{m_\tau^2}{m_B^2}\right )^2\nonumber
\\&&\times f_B^2 |V_{ub}|^2 \tau_B.
\end{eqnarray}
Thus with the theoretical input for B meson decay constant $f_B$,
we can set the bound for $|V_{ub}|$. Figure \ref{fig2} shows the
impact of such bound on the $\bar \rho-\bar\eta$ plane. Combined
with the results of $B_s-\bar B_s$ mixing, the constraints of
$\bar \rho-\bar\eta$ plane is shown in Figure \ref{fig3}. Such
constrained area is consistent with the global fitting (see Figure
\ref{fig3})\cite{CKMfitter}.
\begin{figure}
\epsfig{file=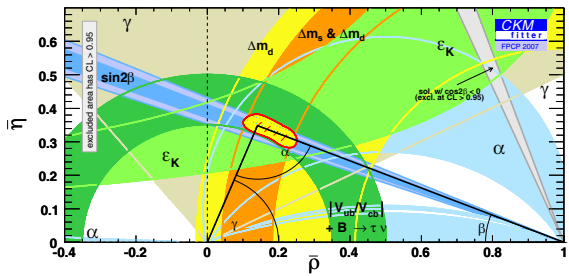,width=8cm,angle=0}
\caption{Global fitting for $\bar \rho-\bar \eta$ plane.}
\label{fig4}
\end{figure}

\begin{figure}
\epsfig{file=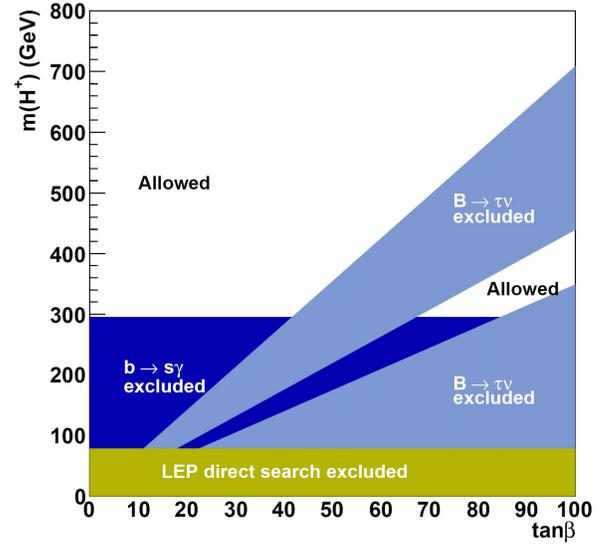,width=8cm,angle=0} \caption{The
constraints of $\tan\beta-m_H$ from $B$ decays.} \label{fig5}
\end{figure}

If we consider the physics beyond the SM, the charged higgs can
also contribute to $B^-\to \tau^-\bar \nu$.
\begin{eqnarray}
\frac{{\rm BR}(B^-\to \tau^-\bar \nu)}{{\rm BR}(B^-\to \tau^-\bar
\nu)_{\rm SM}}&=&\left (1-\frac{m_B^2}{m_H^2}\tan^2\beta\right
)^2.
\end{eqnarray}
This allows us to set the bound the on $\tan\beta/m_H$. Combining
with the results from $B\to X_s\gamma$, the constraints on $\tan
\beta-m_H$ is given in Figure \ref{fig5}\cite{LP2007:BDecays}.

\section{New hadronic resonances at BaBar and Belle}

At BaBar and Belle, many new hadronic resonances are observed. We
list them below.

1) $D^*_{s,J}(2860)$:\\
BaBar measured $e^+e^-\to (DK) X$, and found $DK$ resonance at
mass $m_{DK}\,=\,2856.6\pm1.5\pm5.0$ $\mbox{MeV}/c^2$ with the
decay width $\Gamma\,=\,47\pm 7 \pm 10$ MeV. This could be a
radial excitation of $D^*_{s0}(2317)$.

2) $D_{sJ}(2700)$:\\
In the above analysis, BaBar also found another $DK$ resonance
$X(2690)$ at $ m=(2688\pm4\pm3)$ $\mbox{MeV}/c^2$ with the decay
width $ \Gamma=(112\pm7\pm36)\mbox{MeV}$. Belle confirmed this state
$D_{sJ}(2700)$ in $B^+\to \bar{D}^0 D^0 K^+$ in which $D^0 K^+$
invariant mass peaks at $ m=(2715\pm11^{+11}_{-14})\mbox{MeV}/c^2$
with the decay width $ \Gamma=(115\pm20^{+36}_{-32})\mbox{MeV}$.

3) $X(3940)$, $Y(3940)$ and $Z(3930)$ ($c\bar c$ mesons):\\
Belle observed the $X(3940)$ resonance in
\begin{eqnarray}
e^+e^-\to J/\Psi X(3940):&&
\begin{array}{l}
m=(3943\pm6\pm6)\mbox{MeV}/c^2\\
\Gamma=(15.4\pm 10.1)\mbox{MeV}
\end{array}\nonumber
\end{eqnarray}
$X(3940)$ could be the radially excited charmonium $\eta_c(3s)[3
~^1S_0]$.

Belle also found that in $B\to J/\Psi \omega K$, $J/\Psi\omega$
invariant mass peaks at
\begin{eqnarray}
Y(3940):&&
\begin{array}{l}
m=(3943\pm11\pm13)\mbox{MeV}/c^2\\
\Gamma=(87\pm 22\pm 26)\mbox{MeV}
\end{array}\nonumber
\end{eqnarray}
$Y(3940)$ could be $\chi^\prime_{c1}[2 ~^3P_1]$.

In $\gamma\gamma\to D\bar D$, $D\bar D$ invariant mass peaks at
\begin{eqnarray} Z(3930):&&
\begin{array}{l}
m=(3929\pm5\pm2)\mbox{MeV}/c^2\\
\Gamma=(29\pm 10\pm 2)\mbox{MeV}
\end{array}
\end{eqnarray}\nonumber
$Z(3930)$ could be $\chi^\prime_{c2}[2 ~^3P_2]$.

4) $X(3872)$:\\
Belle found this resonance in
\begin{eqnarray}
B^{\pm}&\to& X(3872) K^\pm \nonumber \\
 &&\hookrightarrow J/\Psi \pi\pi\nonumber
\end{eqnarray}
$J/\Psi \pi\pi$ invariant mass peaks at $m=(3871.81\pm
0.36)\mbox{MeV}/c^2$ with narrow decay width $\Gamma<2.3
\mbox{MeV}$.

5) $Y(4260)$:\\
BaBar observed this resonance in
 $e^+e^-\to\gamma_{\rm isr} (J/\Psi \pi^+\pi^-)$
(isr for initial state radiated).  The $J/\Psi\pi\pi$ invariant
mass peaks at $m=(4295\pm10^{+10}_{-~3})\mbox{MeV}/c^2$ with the
decay width $\Gamma=(88\pm23) \mbox{MeV}$. It is confirmed by
Belle and CLEO.
\begin{eqnarray} &&\mbox{Belle}:
\begin{array}{l}
m=(4295\pm10^{+10}_{-~3})\mbox{MeV}/c^2\\
\Gamma=(133^{+26+13}_{-22-~6})\mbox{MeV}
\end{array}\nonumber \\
&&\mbox{CLEO}: m=(4283^{+17}_{-16}\pm4)\mbox{MeV}/c^2 \nonumber
\end{eqnarray}

\section{$b$-flavored hadron spectrum}
$B$ factories are unable to produce the heavy $b$-flavored hadrons
such as $B_s$, excited $B$ mesons, and $b$-baryons. These hadronic
states can only be accessible at the hadron colliders.

We collect the recent measurements from TEVATRON on the spectra of
$b$-flavored baryons as the following\cite{b-baryon}:
\begin{eqnarray}
m(\Lambda_b)&=&5619.7\pm 1.2~~~\mbox{MeV}\nonumber \\
m(\Sigma_b^{-})&=&5815.2^{+1.0}_{-1.9}\pm1.7~~~\mbox{MeV}\nonumber \\
m(\Sigma_b^{+})&=&5807.5^{+1.9}_{-2.2}\pm1.7~~~\mbox{MeV}\nonumber \\
m(\Sigma_b^{*-})&=&5836.7^{+2.0+1.8}_{-2.3-1.7}~~~\mbox{MeV}\nonumber \\
m(\Sigma_b^{*+})&=&5829.0^{+1.6+1.7}_{-1.7-1.8}~~~\mbox{MeV}\nonumber\\
m(\Xi_b^-)&=&(5774\pm11\pm15)~~~\mbox{MeV} \nonumber
\end{eqnarray}

The status of the excited $L=1$, $j^p=3/2^+$ $B$ mesons ($B_J$,
$B_{sJ}$) masses (MeV) are listed in Table \ref{table4}.
\begin{table}
\begin{tabular}{ccc}\hline\hline
~ & $B^0_1$ & $B^{*0}_2$ \\
\hline
 CDF & $5734\pm3\pm 2$ & $5738\pm 5\pm1$ \\
$D0$ & $5720.8\pm 2.5 \pm 5.3$ & $5746\pm2.4\pm5.4$
 \\\hline\hline
~ & $B_{s1}$ & $B^{*0}_{s2}$ \\\hline
 CDF &  $5829.4\pm
0.2\pm 0.6$ & $ 5839.6\pm 0.4\pm 0.5$ \\
$D0$ & ~ & $5839.1\pm1.4\pm1.5$
\\
\hline\hline
\end{tabular}
\caption{Exited $B$ meson spectra.}\label{table4}
\end{table}

\section{Conclusion}
Almost 30 years after the discovery of $b$ quark, and 7 years $B$
factories running, $B$ physics has entered a precision test era.
The higher order theoretical calculations are essential to explain
the more and more accurate experimental data, especially the data
for non-leptonic decays. This does not only require the
straight-forward but tough computation but also developing new
theoretical concepts in heavy flavor physics.

 For the theoretically clean decays, the new experimental
 measurements shed a light on the precision test of the SM
 and the door towards new physics.

 $B$ factories are also good places to find charmed mesons and
 charmonium states. The recently observed new mesons properties
 still need further theoretical study.

LHC will run in next year, $B$ physics will enter a new era. We
can fully explore all the $b$-flavored hadrons and their decays.
Theorists will find more interesting subjects there.

\section{Acknowledgement}
 \noindent This work is partly supported by the National
Natural Science Foundation of China under grant number 10375073 and
90403024.


\begin{thebibliography}{99}
\bibitem{HFAG}
Heavy Flavor Averaging Group,
\underline{http://www.slac.stanford.edu/xorg/hfag/index.html}.

\bibitem{Beneke:2003zv}
  M.~Beneke and M.~Neubert,
  Nucl.\ Phys.\  B {\bf 675}, 333 (2003)
  [arXiv:hep-ph/0308039].

\bibitem{Yoshikawa:2003hb}
  T.~Yoshikawa,
  Phys.\ Rev.\  D {\bf 68}, 054023 (2003)
  [arXiv:hep-ph/0306147].



\bibitem{Beneke:2005vv}
  M.~Beneke and S.~Jager,
  Nucl.\ Phys.\  B {\bf 751}, 160 (2006)
  [arXiv:hep-ph/0512351].

\bibitem{Beneke:2006mk}
  M.~Beneke and S.~Jager,
  Nucl.\ Phys.\  B {\bf 768}, 51 (2007)
  [arXiv:hep-ph/0610322].

\bibitem{Beneke:2005gs}
  M.~Beneke and D.~Yang,
  Nucl.\ Phys.\  B {\bf 736}, 34 (2006)
  [arXiv:hep-ph/0508250].

\bibitem{Bell:2007tv}
  G.~Bell,
  arXiv:0705.3127 [hep-ph].

\bibitem{Aubert:2003mm}
  B.~Aubert {\it et al.},  
  Phys.\ Rev.\ Lett.\  {\bf 91}, 171802 (2003);
  Phys.\ Rev.\ Lett.\  {\bf 93}, 231804 (2004);
  [hep-ex/0408093];
  K.~F.~Chen {\it et al.}, 
  Phys.\ Rev.\ Lett.\  {\bf 91}, 201801 (2003);
  Phys.\ Rev.\ Lett.\  {\bf 94}, 221804 (2005);
  J.~Zhang {\it et al.},  
  [hep-ex/0505039].



\bibitem{PhiKstar}
  H.~Y.~Cheng and K.~C.~Yang,
  Phys.\ Lett.\  B {\bf 511}, 40 (2001)
  [arXiv:hep-ph/0104090];
  X.~Q.~Li, G.~r.~Lu and Y.~D.~Yang,
  Phys.\ Rev.\  D {\bf 68}, 114015 (2003)
  [Erratum-ibid.\  D {\bf 71}, 019902 (2005)]
  [arXiv:hep-ph/0309136];
  A.~L.~Kagan,
  Phys.\ Lett.\  B {\bf 601}, 151 (2004)
  [arXiv:hep-ph/0405134];
  P.~Colangelo, F.~De Fazio and T.~N.~Pham,
  Phys.\ Lett.\  B {\bf 597}, 291 (2004)
  [arXiv:hep-ph/0406162];
  W.~S.~Hou and M.~Nagashima,
  arXiv:hep-ph/0408007;
  H.~n.~Li and S.~Mishima,
  Phys.\ Rev.\  D {\bf 71}, 054025 (2005)
  [arXiv:hep-ph/0411146];
  Y.~D.~Yang, R.~M.~Wang and G.~R.~Lu,
  Phys.\ Rev.\  D {\bf 72}, 015009 (2005)
  [arXiv:hep-ph/0411211];
  P.~K.~Das and K.~C.~Yang,
  Phys.\ Rev.\  D {\bf 71}, 094002 (2005)
  [arXiv:hep-ph/0412313];
  H.~n.~Li,
  Phys.\ Lett.\  B {\bf 622}, 63 (2005)
  [arXiv:hep-ph/0411305];
  C.~S.~Kim and Y.~D.~Yang,
  arXiv:hep-ph/0412364;
  W.~j.~Zou and Z.~j.~Xiao,
  Phys.\ Rev.\  D {\bf 72}, 094026 (2005)
  [arXiv:hep-ph/0507122];
  C.~S.~Huang, P.~Ko, X.~H.~Wu and Y.~D.~Yang,
  Phys.\ Rev.\  D {\bf 73}, 034026 (2006)
  [arXiv:hep-ph/0511129];
  Q.~Chang, X.~Q.~Li and Y.~D.~Yang,
  JHEP {\bf 0706}, 038 (2007)
  [arXiv:hep-ph/0610280].

\bibitem{RhoKstar}
  H.~W.~Huang, C.~D.~Lu, T.~Morii, Y.~L.~Shen, G.~Song and Jin-Zhu,
  Phys.\ Rev.\  D {\bf 73}, 014011 (2006)
  [arXiv:hep-ph/0508080];
  S.~Baek, A.~Datta, P.~Hamel, O.~F.~Hernandez and D.~London,
  Phys.\ Rev.\  D {\bf 72}, 094008 (2005)
  [arXiv:hep-ph/0508149].

\bibitem{Beneke:2006hg}
  M.~Beneke, J.~Rohrer and D.~Yang,
  Nucl.\ Phys.\  B {\bf 774}, 64 (2007)
  [arXiv:hep-ph/0612290].

\bibitem{Beneke:2005we}
  M.~Beneke, J.~Rohrer and D.~Yang,
  Phys.\ Rev.\ Lett.\  {\bf 96}, 141801 (2006)
  [arXiv:hep-ph/0512258].

\bibitem{Abulencia:2006mq}
  A.~Abulencia {\it et al.}  [CDF - Run II Collaboration],
  Phys.\ Rev.\ Lett.\  {\bf 97}, 062003 (2006)
  [AIP Conf.\ Proc.\  {\bf 870}, 116 (2006)]
  [arXiv:hep-ex/0606027];
  V.~M.~Abazov {\it et al.}  [D0 Collaboration],
  Phys.\ Rev.\ Lett.\  {\bf 97}, 021802 (2006)
  [arXiv:hep-ex/0603029].

\bibitem{Ikado:2006un}
  K.~Ikado {\it et al.},
  Phys.\ Rev.\ Lett.\  {\bf 97}, 251802 (2006)
  [arXiv:hep-ex/0604018].

\bibitem{:2007xj}
  G.~Nardo {\it et al.}  [BABAR Collaboration],
  arXiv:0708.2260 [hep-ex].

  \bibitem{CKMfitter}
  CKMfitter Group, \underline{http://ckmfitter.in2p3.fr/}.

\bibitem{LP2007:BDecays}
M. Nakao, talk given at LP2007, Korea.

\bibitem{b-baryon}
CDF colaborations, hep-ex/0702047; D0 colaborations,
hep-ex/0706.1690.


\end{thebibliography}
\end{document}